\documentclass[twocolumn,showpacs,preprintnumbers,amsmath,amssymb]{revtex4}

\usepackage{graphicx}
\usepackage{dcolumn}
\usepackage{bm}

\begin{document}

\preprint{APS/123-QED}

\title{Reversing the training effect in exchange biased CoO/Co bilayers}

\author{Steven Brems}
\email{steven.brems@fys.kuleuven.ac.be}
\author{Dieter Buntinx}
\author{Kristiaan Temst}
\author{Chris Van Haesendonck}
\affiliation{Laboratorium voor Vaste-Stoffysica en Magnetisme, Katholieke Universiteit Leuven, Celestijnenlaan 200
D, B-3001 Leuven, Belgium}

\author{Florin Radu}
\author{Hartmut Zabel}
\affiliation{ Experimentalphysik/Festk\"{o}rperphysik, Ruhr-Universit\"{a}t Bochum, D-44780 Bochum, Germany}

\date{\today}

\begin{abstract}
We performed a detailed study of the training effect in exchange biased CoO/Co bilayers.
High-resolution measurements of the anisotropic magnetoresistance (AMR) are consistent with
nucleation of magnetic domains in the antiferromagnetic CoO layer during the first magnetization
reversal. This accounts for the enhanced spin rotation observed in the ferromagnetic Co layer for
all subsequent reversals. Surprisingly, the AMR measurements as well as magnetization measurements
reveal that it is possible to partially reinduce the untrained state by performing a hysteresis
measurement with an in plane external field perpendicular to the cooling field. Indeed, the next
hysteresis loop obtained in a field parallel to the cooling field resembles the initial asymmetric
hysteresis loop, but with a reduced amount of spin rotation occurring at the first coercive field.
This implies that the antiferromagnetic domains, which are created during the first reversal after
cooling, can be partially erased.
\end{abstract}


\maketitle

The exchange bias (EB) effect is observed when a layer of a ferromagnet (FM) makes contact with a
layer of an antiferromagnet (AF), which introduces an exchange coupling at their interface. This
results in a unidirectional shift of the hysteresis loop when the bilayer is grown in a magnetic
field or cooled in a field below the N\'eel temperature ($T_N$) of the AF. The EB in the AF/FM
bilayers also gives rise to an enhanced coercivity as well as to an asymmetric reversal of the
magnetization, which can be strongly affected by ``training", i.e., by going through consecutive
hysteresis loops. The EB, which was recently linked to a fraction of uncompensated interfacial
spins (about 4 to 7\% of a monolayer) that are pinned to the AF and are not affected by an
external field~\cite{1Ohldag,2Kappenberger}, was discovered almost 50 years ago by Meiklejohn and
Bean~\cite{3Meiklejohn}. A reliable theoretical understanding is however still
lacking~\cite{4Nogues,5Berkowitz,6Stamps,7Kiwi}. Therefore, and because of technological
applications such as spin valves in magnetic reading heads and magnetic random access memories,
the EB effect remains at the forefront of research in thin film magnetism.

In this letter, we report on the results of a detailed study of the training effect in
CoO(AF)/Co(FM) bilayers. Polycrystalline CoO/Co bilayers are selected due to their very pronounced
training effects: the coercivity decreases and the shape of the magnetization loop changes
considerably. Several theoretical models have been put forward to explain the training effect, but
a detailed understanding of the effect is missing. The domain state model, which states that the
EB shift results from an exchange field provided by irreversible magnetization of the AF, enables
to explain the training effect in terms of domain wall formation \emph{perpendicular to the
interface} in the AF ~\cite{15Nowak,16Keller}. When going through the hysteresis loop, a
rearrangement of the AF domain structure results in a partial loss of the domain state
magnetization and causes a reduction of the EB effect. Irreversible training effects can also be
related to the symmetry of the antiferromagnetic anisotropies and the inherent frustration of the
interface~\cite{17Hoffmann}. Radu et al.~\cite{10Radu} argued that the asymmetry is caused by
interfacial domain formation (\emph{parallel to the interface}) during the very first reversal.
These interfacial domains serve as seeds for the subsequent magnetization reversals. Here, we show
that the untrained state can be re-induced by going through an hysteresis loop with the applied
magnetic field perpendicular to the cooling field direction without raising the temperature above
$T_N$. This surprising effect is directly reflected by magnetization measurements performed with a
superconducting quantum interference device (SQUID) magnetometer. High-resolution measurements of
the magnetoresistance allow us to further elucidate this partial reversibility of the training
effect.

In a FM layer the resistance depends on the angle between the magnetization and the current
direction. This angle-dependent resistance is known as the anisotropic magnetoresistance
(AMR)~\cite{8Campbell,9McGuire}. In a saturated FM layer, the AMR effect can be expressed as
\begin{equation}
R(\theta)=R_{\bot}+\triangle R_o \cos^2(\theta) \; , \label{magnetoresistance}
\end{equation}
where $R_{\bot}$ is the resistance with the magnetization perpendicular to the current and
$\triangle R_o$ is the difference in resistance with the magnetization parallel and perpendicular
to the current, respectively. The origin of the AMR effect is related to spin-orbit scattering.
For the present study AMR measurements are performed to probe in detail the switching behavior of
the CoO/Co bilayers for different subsequent hysteresis loops.

\begin{figure}
\includegraphics{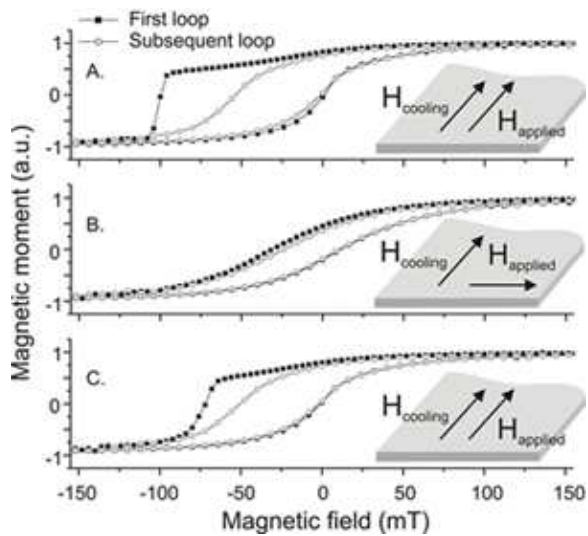}
\caption{\label{radu} SQUID magnetization measurements of a CoO/Co bilayer at 10\,K after cooling
in a field of +100\,mT. The upper panel (a) shows the first and second hysteresis loop with the
magnetic field applied in the direction of the cooling field. Panel (b) represents the subsequent
two hysteresis loops when the magnetic field is applied perpendicular to the cooling field. The
lower panel (c) shows the next two hysteresis loops with the magnetic field again applied along
the cooling field direction. A re-entry of the untrained state without heating the sample above
the blocking temperature is observed.}
\end{figure}

For the preparation of the CoO/Co bilayers a 20\,nm thick Co layer is dc magnetron sputtered on
top of an oxized Si wafer with a typical deposition rate of 0.1\,nm/s. The base pressure of the
vacuum sputter chamber is $10^{-7}$\,mbar, while the working pressure for the Ar sputter gas is
$10^{-3}$\,mbar. After deposition, the Co layer is oxidized \emph{in-situ} for 2 minutes in a
partial oxygen pressure of $10^{-3}$\,mbar, which results in the formation of a 2\,nm thick CoO
top layer. For the SQUID magnetization measurements the sample is cooled to 10\,K, which is well
below the blocking temperature, in a field of +100\,mT in the sample plane. After field cooling,
the magnetic field is increased to +200\,mT and two subsequent hysteresis loops (Fig.\,1(a)) are
measured with the field parallel to the cooling field. The first reversal at -100\,mT is more
abrupt, while all subsequent reversals are more rounded. This asymmetric behavior is typical for
the training effect in CoO/Co and can be directly linked to a change in the magnetization reversal
mechanism. Initially, domain wall nucleation and domain wall propagation govern the reversal,
leading to a sudden change of the magnetization. The following more rounded reversals are
dominated by a rotation of the magnetization~\cite{10Radu,11Gierlings}. This training effect can
be understood as being the result of the splintering of the AF into a collage of domains during
the first reversal at negative fields~\cite{12Welp}. Throughout field cooling the ferromagnetic Co
layer consists of a single FM domain, which induces a uniform state in the AF CoO. During the
first reversal, the uniform FM Co magnetization is broken up and via the exchange coupling at the
CoO/Co interface this results in a torque acting on the CoO spins. As a result, the metastable
uniform AF state lowers its interfacial energy by splitting up into domains. The latter AF domain
structure will affect all subsequent magnetization reversals~\cite{18Nolting,19Ohldag}. Figure
1(b) shows the subsequent two SQUID magnetization measurements of the hysteresis loop with the
magnetic field perpendicular to the cooling field. Almost no EB or training effect is observed.
Finally, when the external magnetic field is again applied along the cooling field direction, we
surprisingly observe the reappearance of an asymmetric hysteresis loop. Remarkably, the untrained
state can be partially reinduced by changing the orientation of the applied magnetic field and
this without heating the sample above the N\'{e}el temperature.

\begin{figure}
\includegraphics{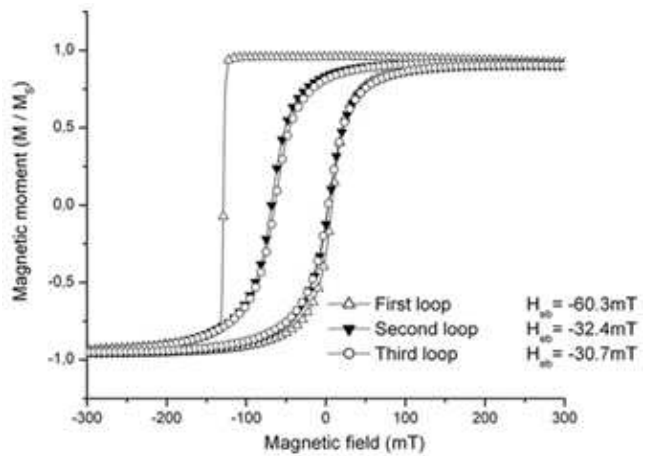}
\caption{\label{vsm} Hysteresis loops measured at 5\,K with VSM magnetometry of a CoO/Co bilayer
cooled in a field of +400\,mT. The first reversal at negative field is dominated by domain wall
nucleation and domain wall propagation and is abrupt. All subsequent reversals are dominated by
rotation of the magnetization and are more rounded.}
\end{figure}

To further elucidate the partial reappearance of the untrained state, measurements of the AMR were
performed. The AMR provides direct information about the domain configuration of the FM and, as a
result of the pinning also about the AF. For the high-resolution magnetoresistance measurements we
fabricate narrow stripes of CoO/Co using electron-beam lithography and lift-off techniques. After
exposure and development of the resist layer, a CoO(2\,nm)/Co(20\,nm) bilayer is deposited by
sputtering and subsequent in-situ oxidation. Finally, the lift-off is performed by immersing the
sample in a bath of hot acetone. In order to increase the sensitivity of our magnetoresistance
measurement, $2\,\mu$m wide and $120\,\mu$m long stripes are fabricated. Both ends of a stripe are
connected to larger predefined Au contact pads to which we are able to attach the voltage and
current leads by ultrasonic wire bonding. High-resolution four-terminal magnetoresistance
measurements are performed in a helium flow cryostat by integrating the sample into an
Adler-Jackson bridge. The ac measuring current for the lock-in detection has a frequency of
27.7\,Hz and a root-mean-square (rms) amplitude of $3.5\,\mu$A.

The results of our magnetization measurements with a vibrating sample magnetometer (VSM) on an
unpatterned CoO/Co reference film, which is deposited simultaneously with a CoO/Co stripe, are
shown in Fig.~\ref{vsm}. The sample is cooled to 5\,K in an in-plane field of +400\,mT. The first
reversal in the decreasing field branch at $\rm{-130}$\,mT is very abrupt while all subsequent
reversals are more rounded, in agreement with the results obtained with SQUID magnetometry for the
CoO/Co sample discussed above.

\begin{figure}
\includegraphics{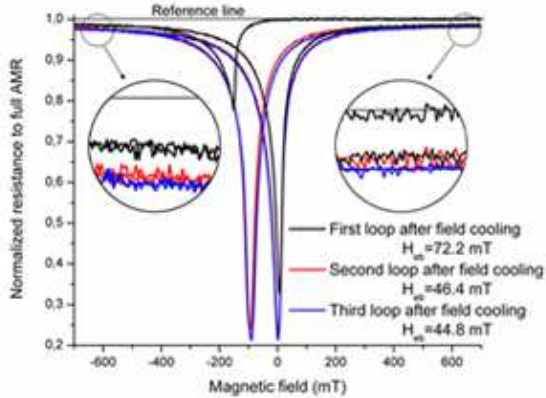}
\caption{\label{training} Field dependence of the magnetoresistance of a CoO/Co stripe at\,10 K
after cooling in a field of +100\,mT applied along the length of the stripe. A smaller resistance
change (less rotation) is observed during the first reversal when compared to the subsequent
reversals. The insets compare the resistance at saturation to the maximum resistance (reference
line), which ideally corresponds to the case that all spins are oriented along the cooling field
direction.}
\end{figure}

Figure \ref{training} shows the magnetoresistance measurements of the CoO/Co stripe after cooling
to 10\,K in a field of +100\,mT parallel to the stripe. After field cooling, the magnetic field is
increased to +700\,mT and three subsequent hysteresis loops are measured with the field parallel
to the CoO/Co stripe. A smaller AMR effect (less rotation) is observed for the first reversal when
compared to the subsequent reversals. These AMR results are consistent with our VSM magnetometry
(see Fig.~\ref{vsm}) as well as with previous results~\cite{13Gredig,14Gruyters}. More interesting
is the direct indication for the existence of magnetic domains in the Co layer. After field
cooling and before passing through the first magnetization reversal in the descending field
branch, the resistance in saturation reaches its maximum because all spins are oriented along the
cooling field. After going through a complete hysteresis loop, the resistance in saturation is
reduced (see right inset in Fig.~\ref{training}), indicating that the spins in the FM are canted
away from the cooling field, which is consistent with a domain structure present in the FM. These
domains originate from the AF, which is strongly coupled to the FM by the exchange interaction.
Therefore, our AMR results are consistent with the fact that the AF splits up into domains after
the first reversal. As reported before~\cite{13Gredig}, the training effect in CoO/Co bilayers
depends on the thickness of the AF layer. Bilayers with thicker CoO (thickness larger than 5\,nm)
reveal less training and relatively square hysteresis loops. In thinner CoO layers (thickness
smaller than 5\,nm) similar to our CoO layers, changes in the spin alignment of the AF grains are
possible because of their smaller magnetocrystalline anisotropy. As revealed by our measurements,
the training effect in this type of films is consistent with the altering of the CoO spin
structure. Quantitatively, the resistance in saturation is reduced by 1,6\,\% after going through
a complete hysteresis loop (inset Fig.~\ref{training}). Using Eq.~\ref{magnetoresistance} we find
that such a reduction is consistent with the formation of domain walls parallel to the AF/FM
interface, where the domain walls extend over a few monolayers~\cite{10Radu}.

\begin{figure}
\includegraphics{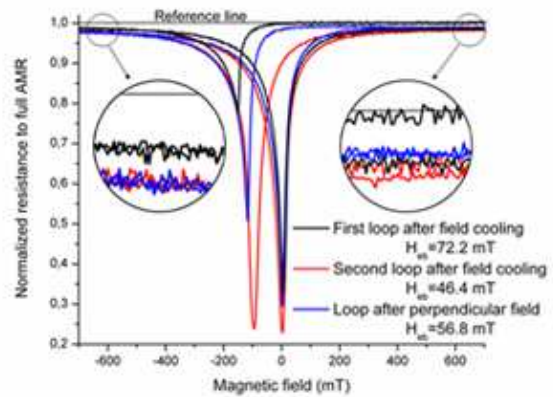}
\caption{\label{training_perp} Field dependence of the magnetoresistance of the CoO/Co stripe at
10\,K after cooling in a field of +100\,mT along the stripe. The blue line illustrates the
reappearance of the training effect without any heating of the sample. This reappearance is
achieved by going through a hysteresis loop with the magnetic field in the sample plane but
perpendicular to the cooling field direction (not shown). The insets show the resistance at
saturation when compared to the maximum resistance (reference line).}
\end{figure}

Our magnetoresistance measurements confirm that it is possible to partially reinduce the untrained
state without heating the sample above the N\'{e}el temperature. This implies that the magnetic
state obtained after field cooling is less irreversible and unique than generally accepted.
Figure~\ref{training_perp} shows two hysteresis loops along the cooling field direction after
field cooling to 10\,K in a field of +100\,mT.
After going through several hysteresis loops, a reversed training effect can be achieved by going
through a hysteresis loop with the magnetic field in the sample plane but perpendicular to the
cooling field direction (not shown). After performing the loop in the perpendicular field, a
hysteresis loop is measured with the field again applied along the cooling field direction. It is
clear from Fig.~\ref{training_perp} that the untrained state has been partially reinduced without
any heating of the sample. The exchange bias field is increased and the amount of magnetization
rotation in the descending field branch is reduced when compared to the trained reversals. An
indication for the mechanism governing this partial reappearance of the untrained state can be
obtained from the magnetoresistance at saturation (see right inset in Fig.~\ref{training_perp}).
After performing the hysteresis loop in the perpendicular field, the initial magnetoresistance at
saturation is again higher than the magnetoresistance after the trained reversal. From these
results we conclude that performing a hysteresis loop in a field perpendicular to the cooling
field alters or partially removes the FM domains. Because the AF domains, which are coupled by a
fraction of uncompensated interfacial spins~\cite{1Ohldag,2Kappenberger} to the FM, are inducing
the FM domains, it is very likely that the domain structure of the CoO is also altered by the
application of the perpendicular field. When performing a hysteresis loop in a perpendicular field
for the second time, we observe a similar behavior although the partial revival of the untrained
state is less pronounced when compared to the revival after the first loop in a perpendicular
field. A more detailed analysis of our results~\cite{20radu} indicates that the external field not
only affects the AF domain size distribution, but also induces a collective rotation of the AF
spins.

In conclusion, the results of our magnetization and magnetoresistance experiments demonstrate that
it is possible to partially reinduce the untrained state in an exchange biased CoO/Co structure. A
clear increase in exchange bias field and a reduction in the amount of magnetization rotation is
observed after performing a hysteresis loop in a magnetic field perpendicular to the cooling field
direction. This surprising result can be explained by a change in the magnetic domain structure in
the antiferromagnetic CoO layer by the application of the perpendicular field. The presence of
antiferromagnetic domains is confirmed by a careful inspection of the magnetoresistance data at
saturation.

This work has been supported by the Fund for Scientific Research - Flanders (FWO) as well as by
the Flemish Concerted Action (GOA) and the Belgian Interuniversity Attraction Poles (IAP) research
programs. F.R. and H.Z. acknowledge support through SFB 491 of the Deutsche
Forschungsgemeinschaft.

\newpage
\bibliography{other-prl-brems}

\end{document}